# State-Machine Governance of Adaptive Enforcement: Evidence from a Paired-Scenario Evaluation

Ifeanyi Williams AGBEDO and Yanzhen QU
Colorado Technical University
Colorado Springs, CO, USA
williams.agbedo@student.ctuonline.edu; yqu@coloradotech.edu

**Abstract: AI-assisted security systems can recommend rapid policy changes, but governing their execution requires explicit authorization, evidence, and recovery rules. This study evaluated a machine-verifiable, state-machine-governed transition procedure implemented in ZTGuard using 36 matched scripted scenarios and deterministic vendor-connector simulations. Each pair held the starting state, telemetry, AI recommendation, and scheduled vendor condition constant. The sole dependent variable, Control, was the equal-weight mean of traceability, accountability, audit-chain completeness, and reliable rollback capability. Mean Control was 0.720 (SD = 0.218) for the scripted operator-driven baseline and 0.972 (SD = 0.131) for the governed procedure. A two-tailed paired Wilcoxon signed-rank test yielded V = 359, p < .001, with a matched-pairs rank-biserial correlation of .899. Twenty-six pairs favored treatment, nine were tied, and one favored baseline. The findings support controlled artifact utility, not production readiness. Interpretation is limited by purposive sampling, scripted operators, simulated connectors, overlapping components, and arm-specific scoring rules.**

## 1. Introduction

Zero Trust Architecture replaces location-based assumptions with continuous verification, least privilege, and explicit policy enforcement [1]. Cloud-native guidance extends those principles to distributed policy decision and enforcement components across multi-location environments [2]. Yet these architectures provide comparatively limited direction for the operational interval after an AI-assisted system recommends a policy change and before that change is committed across heterogeneous enforcement platforms. In that interval, organizations must determine whether a transition is authorized, which evidence supports it, whether dependent enforcement points converge, and how the previous validated state will be restored if the transition fails.

The problem becomes consequential when adaptive enforcement operates faster than informal review and vendor-specific workflows can consistently support. AI-driven cybersecurity can accelerate analysis and response, but it also creates a need for transparent and accountable governance of machine-assisted decisions [3]. Without a formal transition layer, movement between operational states can depend on variable operator practices and disconnected platform logs. A policy may reach the intended enforcement posture while leaving incomplete evidence about who or what initiated the change, why it was allowed, whether all enforcement points reached the same state, or whether recovery remained possible.

This article evaluates whether a machine-verifiable, state-machine-governed transition procedure improves measured Control relative to a scripted operator-driven procedure under matched inputs. The procedure models Monitor, Alert, and Enforce; block, deny, and quarantine are actions within Enforce, not additional states. Machine-verifiability describes automatically checkable transition conditions and execution evidence, not an additional hypothesis-tested outcome or an exhaustive mathematical proof of system correctness. Control is the equal-weight composite of traceability, accountability, audit-chain completeness, and reliable rollback capability.

The quantitative Design Science Research evaluation executed 36 scenarios once per condition using CICIDS2017, CICIoT2023, and explicitly labeled synthetic telemetry. Deterministic simulations represented Check Point Smart-1 and Palo Alto Panorama connectors.

## 2. Related Work

### 2.1. Zero Trust and the Adaptive-Enforcement Transition Gap

NIST Zero Trust Architecture separates policy decision, administration, and enforcement responsibilities and requires explicit verification before access is granted [1]. NIST guidance for cloud-native applications further addresses distributed controls in multi-location environments [2]. These models establish essential architectural responsibilities, but they do not fully prescribe a vendor-neutral lifecycle for moving an AI-informed policy among operational states while preserving transition evidence and recovery requirements. The unresolved issue is not whether access should be verified; it is how a recommendation becomes an attributable, auditable, and reversible operational change.

AI-driven cybersecurity research addresses detection, intelligence, and automated response [3]. The present study focuses instead on the transition from recommendation to enforcement: coordinating tools does not, by itself, establish authorization, completion, or rollback requirements.

### 2.2. Finite-State and Formal Transition Governance

Finite-state machines provide an explicit representation of system states, permitted paths, and observable outputs. Their testing literature shows how specified state movement can support systematic evaluation of conformance [4]. Formal-methods scholarship further distinguishes an explicit behavioral specification from evidence that an implementation satisfies that specification [5]. Applied to adaptive enforcement, these ideas support representing Monitor, Alert, and Enforce as discrete states and allowing movement only when specified guards, preconditions, and postconditions are satisfied.

The artifact implemented these principles in a typed transition kernel that returned a defined post-state or typed failure. AI recommendations could not independently commit policy changes. This separated recommendation from enforcement authority and made transition requirements inspectable without claiming exhaustive formal verification.

### 2.3. Accountability, Provenance, and Audit Evidence

Accountability requires evidence that connects an action with the responsible entity and the basis for the action [6]. Secure logging and information-accountability research likewise emphasize tamper-evident histories, attribution, and the ability to reconstruct consequential system behavior [7], [8]. Provenance scholarship frames this need as preserving the origins and transformations necessary to explain how an outcome was produced [9]. These ideas are directly relevant to policy transitions because a final platform state alone does not explain the actor, supporting evidence, prior state, transition result, or sequence of events.

The evaluated harness used an in-memory, sequenced audit chain containing actor identity, an evidence pointer, pre-state, observed post-state, and sequence information. Both conditions used a seven-field record schema: timestamp, actor, policy identifier, pre-state, post-state, evidence pointer, and outcome. Shared fields did not imply identical scoring predicates. Cryptographic chain verification was reported separately but also served as a prerequisite for treatment traceability; it was not a fifth component or a direct term in the four-component mean.

### 2.4. Feedback, Convergence, and Rollback

Feedback-loop research describes self-adaptive systems as monitoring observed behavior, comparing it with objectives, and initiating corrective action when behavior diverges [10], [11]. At the policy-transition layer, this logic requires more than initiating a change. The procedure must verify convergence across dependent enforcement points and initiate a bounded return to a previously validated state when the postcondition is not satisfied. Rollback-recovery scholarship similarly treats recovery as a protocol with preserved state and explicit assumptions about failure behavior rather than as a guarantee that every external dependency will recover on demand [12].

The artifact therefore defined rollback as a governed transition. Each applicable path identified the prior validated state, corrective action, and completion bound. This design connected reliable rollback capability to Control while retaining an important boundary: formal recovery logic can detect and expose delayed external-platform behavior, but it cannot force an impaired platform to respond within the required time.

### 1.2.5. Research Gap and Contribution

The contribution is an evaluated integration of explicit transition guards, attributable evidence, post-action convergence checks, and bounded rollback. Its technical value is the checkable relationship among these mechanisms at the recommendation-to-enforcement boundary, including detectable refusal-evidence and recovery failures. Table I contrasts this integration with the scope of cited literature; it does not establish that competing products lack these capabilities. No first-of-kind priority or superiority over SOAR is claimed.

## 3. Problem, Purpose, Research Question and Hypotheses

### 3.1. Problem Statement

The problem is that policy transitions from AI-generated recommendations to operational enforcement in multi-cloud Zero Trust environments may not consistently produce reconstruct able, attributable, auditable, and recoverable transition records because existing transition processes may lack explicit governance mechanisms such as formal state-transition rules, authorization guards, evidence requirements, convergence verification, and bounded rollback procedures.

Table I. Positioning Relative to the Cited Literature

| Research focus | Relationship to this study |
|---|---|
| Zero Trust [1], [2] | Defines access and enforcement responsibilities; this study evaluates transition-level governance. |
| FSM/formal methods [4], [5] | Supplies state and verification foundations; this study does not claim new FSM theory or exhaustive proof. |
| Evidence [6]–[9] | Supplies accountability, logging, and provenance foundations integrated into transition records. |
| Adaptation/recovery [10]–[12] | Supplies feedback and recovery concepts combined with the transition procedure. |
| Evaluated contribution | Matched evidence for the integrated procedure versus a scripted baseline, under stated limits. |

### 3.2. Purpose Statement

The purpose of this quantitative Design Science Research study is to evaluate whether a machine-verifiable, state-machine-governed transition procedure improves Control, measured as the composite of traceability, accountability, audit-chain completeness, and reliable rollback capability, compared with a scripted operator-driven procedure under matched multi-cloud Zero Trust scenarios.

### 3.3. Research Question

Does a machine-verifiable, state-machine-governed transition procedure produce a significant difference in the composite Control Score compared with a scripted operator-driven procedure under matched multi-cloud Zero Trust scenarios?

### 3.4. Hypotheses

**H0:** There is no statistically significant difference in composite Control Score between the machine-verifiable, state-machine-governed procedure and the scripted operator-driven procedure under the matched scenarios.

**H1:** There is a statistically significant difference in composite Control Score between the two procedures under the matched scenarios. Improvement under the governed procedure was expected, but the hypothesis test was two-tailed.

## 4. Methodology

### 4.1. Method

The study used quantitative Design Science Research with a controlled, paired evaluation of a method artifact [13], [14]. The comparison formed part of a build-and-evaluate cycle [15]. A controlled artificial setting was appropriate for evaluating an artifact whose erroneous enforcement action could create operational risk [16].

The independent variable was procedure type: a scripted operator-driven baseline and the machine-verifiable, state-machine-governed treatment. No live operators participated; deterministic cautious, hasty, and nominal profiles modeled baseline variation. Each pair held starting state, telemetry, recommendation, and scheduled vendor conditions fixed. The conditions differed in procedure and associated arm-specific scoring predicates. Matching therefore did not isolate governance behavior from measurement-rule differences; the comparison concerns the two operationalized conditions.

The artifact was instantiated in ZTGuard as a typed transition kernel with explicit guards, preconditions, postconditions, convergence verification, and bounded rollback. ZTGuard was the decision-plane environment; the scientific artifact was the transition procedure. Vendor-neutral interfaces resolved governed intent through deterministic simulations of Check Point Smart-1 and Palo Alto Panorama connectors. The study did not test production vendor control planes, AI recommendation accuracy, or any later application capabilities.

Figure 1 locates the governed procedure between fixed scenario inputs and simulated enforcement. Observed convergence is checked after simulated enforcement, not before it. The diagram describes the evaluated treatment, not a production deployment or a separately tested AI recommendation engine.

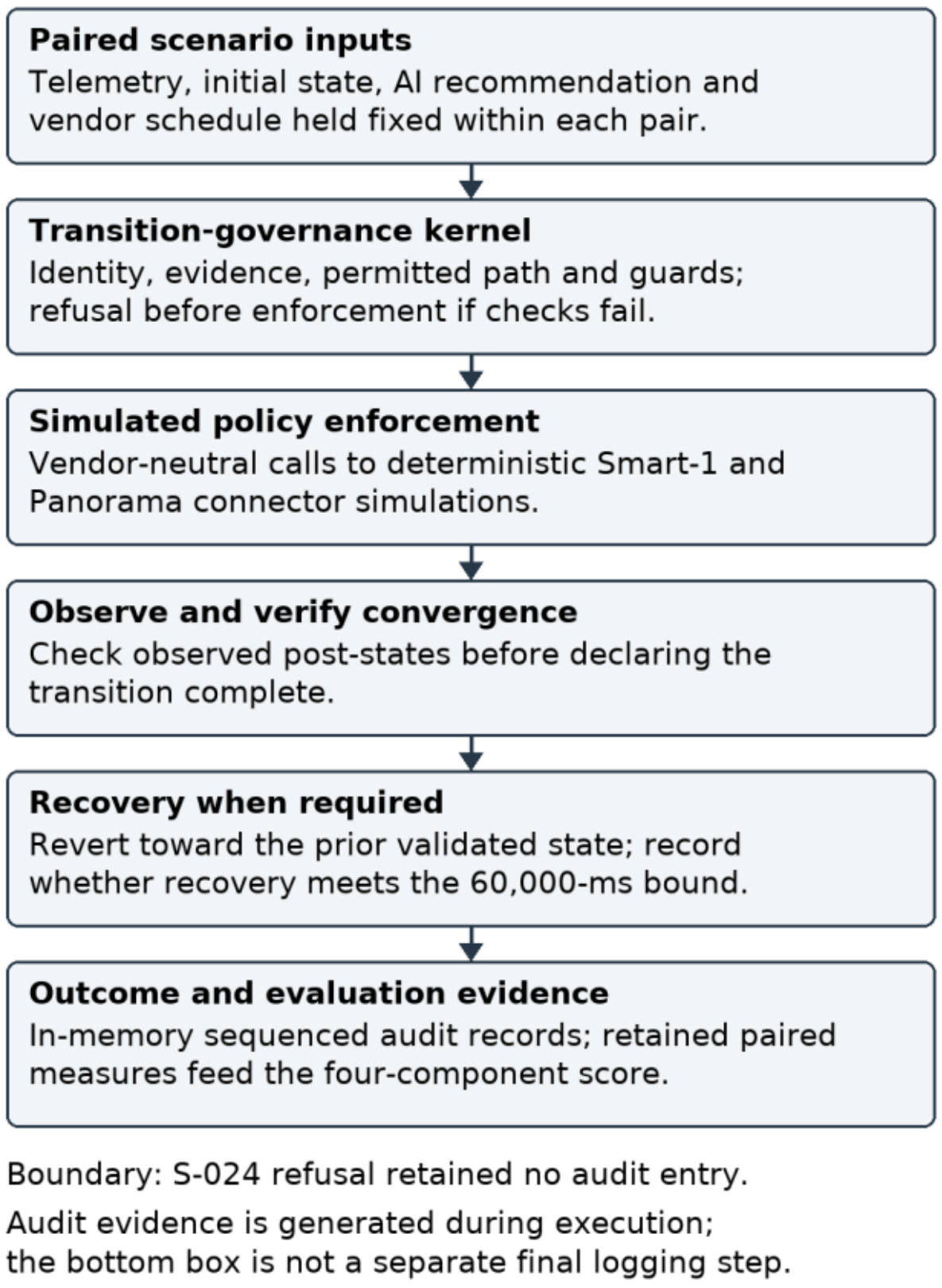

Figure 1. Evaluated transition-governance architecture and evidence boundary.

Control ranged from 0 to 1 and was the equal-weight mean of traceability, accountability, audit-chain completeness, and reliable rollback capability. Equal weighting assigned each requirement one quarter of the composite; it did not establish equal operational importance or empirical independence. The scoring rules were as follows: Traceability: treatment required a last record with the scenario's starting state and a valid audit chain; baseline required any log entry with both states. Accountability: treatment required a nonempty actor and evidence pointer on the last record; baseline checked any entry for a nonempty actor and a defined evidence pointer. Audit-chain completeness: both conditions used the common seven fields, but treatment scored the last record while baseline averaged across entries; some presence tests also differed. Reliable rollback capability: required reversions had to complete within 60,000 ms in both conditions. Cases requiring no reversion received a score of 1. Safety and consistency were descriptive context, not additional hypothesis-tested outcomes.

The Colorado Technical University Institutional Review Board recorded an Exempt and Approved determination dated July 20, 2026. The evaluation used public secondary intrusion-detection data, synthetic scenarios, and scripted operator profiles rather than recruited human participants. Ethical treatment followed the principles applicable to secondary data and information and communication technology research.

#### 4.1.1. Development of the Control Construct

The four components operationalized reconstruction, attribution, evidence completeness, and recovery, drawing on accountability, provenance, secure-logging, and rollback scholarship [6]– [9], [12] (Table II). Equal weighting provided a transparent study-specific definition without an empirically validated basis for differential weights. It was not prescribed by these sources. Detection accuracy, latency, availability, and human usability were outside the primary governance measure rather than unimportant. The arithmetic mean allows compensation across components, so individual failures must remain visible.

Table II. Study-Specific Control Components

| Component | Governance purpose |
|---|---|
| Traceability | Reconstruct recorded state movement. |
| Accountability | Attribute actions to actors and supporting evidence. |
| Audit-chain completeness | Assess presence of the common evidence fields; distinct from cryptographic validity. |
| Reliable rollback | Assess required recovery against its bound; no-reversion cases score 1. |

### 4.2. Population and Sample

The target population comprised policy-state transitions in AI-driven adaptive enforcement across multi-cloud Zero Trust environments. Because it could not be enumerated for probability sampling, 36 scripted scenarios were purposively selected to exercise the Monitor–Alert–Enforce lifecycle and vendor-connector simulations.

The library deliberately covered canonical transitions, guard failures, rollback paths, and convergence conditions. Eleven scenarios used CICIDS2017 telemetry [17], 11 used CICIoT2023 telemetry [18], and 14 used synthetic telemetry for conditions not available directly from the intrusion datasets, such as unauthenticated attempts, connector failures, and rollback-bound breaches. All 36 planned pairs were retained; none was missing or excluded.

Scenario construction used fixed source/category slots and sequential matching records from sorted extracts, rather than a random sample. Label-derived confidence values supplied fixed recommendations: attack records mapped to 0.85–0.95 and benign records to 0.15–0.45. These were deterministic inputs, not predictions from an evaluated learning model. The confidence threshold was 0.7, and the shared rollback bound was 60,000 ms. Explicit synthetic cases exercised missing actions, absent authentication, connector faults, and bound breaches. Reproduction requires retaining the selected rows, scenario paths, operator profiles, and scheduled faults; substituting fresh telemetry would define a different evaluation.

TABLE III. COMPOSITION OF THE PAIRED SCENARIO LIBRARY

| Scenario category | CICIDS2017 | CICIoT2023 | Synthetic | Total |
|---|---|---|---|---|
| Canonical transition | 6 | 6 | 2 | 14 |
| Guard failure | 3 | 3 | 4 | 10 |
| Rollback | 2 | 2 | 4 | 8 |
| Convergence | 0 | 0 | 4 | 4 |
| Total | 11 | 11 | 14 | 36 |

Executing every scenario under both conditions yielded 36 matched pairs and 72 condition-level observations. The baseline used 13 cautious, 12 hasty, and 11 nominal scripted profiles, not observed human behavior.

### 4.3. Data Analysis

The dataset was screened for complete matched records. Descriptive statistics were calculated for the composite score and its four components. Paired Control-Score differences were assessed for normality using the Shapiro–Wilk test. Because normality was not supported, the primary two-tailed hypothesis test used the paired Wilcoxon signed-rank procedure at $\alpha = .05$. The retained analysis used R 4.4.3 with continuity correction and confidence intervals. Nine zero differences were not ranked, leaving 27 nonzero differences for the signed-rank calculation.

The analysis also reported a Hodges–Lehmann location-shift estimate, its 95% confidence interval, and matched-pairs rank-biserial correlation. Component means were descriptive, not separate hypothesis tests. All unfavorable and limit-testing cases were retained.

Matching held scenario inputs constant, supporting within-library comparability. Shared record fields and four named components made scoring inspectable but did not establish identical measurement between conditions. Consequently, the observed Control difference may reflect both procedural behavior and scoring design; its magnitude cannot be attributed solely to governance mechanisms. Component overlap, purposive sampling, scripted operators, and simulated connectors further limit interpretation [19].

### 4.4. Exploratory Retained-Data Analysis

After feedback, a separate analysis loaded the retained 36-pair JSONL file without re-executing or changing scenarios; its SHA-256 matched the recorded manifest. Component scores were reweighted across all 286 nonnegative weight vectors summing to 1 in steps of 0.1, including single-component extremes. Four additional contrasts assigned 0.4 to one component and 0.2 to each other component. A paired percentile bootstrap resampled 36 whole scenario pairs with replacement 20,000 times (Python standard library; seed 86504), using the stored treatment-minus-baseline Control differences and linearly interpolated 2.5th/97.5th percentiles. Rounded retained values were used; original inference and scoring were not replaced.

## 5. Results

### 5.1. Results

The operator-driven baseline produced a mean Control Score of 0.720 (SD = 0.218), while the state-machine-governed procedure produced a mean of 0.972 (SD = 0.131). The observed mean paired improvement was 0.252 points. The median increased from 0.714 to 1.000. Twenty-six pairs had a higher treatment score, nine were tied, and one had a lower treatment score.

TABLE IV. COMPOSITE CONTROL SCORE AND PRIMARY PAIRED ANALYSIS

| Measure | Operator-driven baseline | State-machine treatment | Paired result |
|---|---|---|---|
| Mean Control Score | 0.720 | 0.972 | Mean difference = 0.252 |
| Standard deviation | 0.218 | 0.131 | SD of differences = 0.253 |
| Median | 0.714 | 1.000 | Observed median difference = 0.286 |
| Minimum | 0.357 | 0.250 | Minimum difference = -0.464 |
| Maximum | 1.000 | 1.000 | Maximum difference = 0.643 |
| Shapiro–Wilk test | — | — | $W = 0.891$, $p = .002$ |
| Wilcoxon signed-rank test | — | — | $V = 359$, $p < .001$ |
| Hodges–Lehmann shift | — | — | 0.321, 95% CI [0.295, 0.447] |
| Rank-biserial effect | — | — | $r = .899$ |

The retained 36-pair data reproduced the primary R 4.4.3 analysis: $V = 359$, two-tailed $p = 0.000042782$, and rank-biserial $r = 0.899471$. Table IV reports rounded values. The Hodges–Lehmann estimate, 0.321 (95% CI [0.295, 0.447]), is a paired location-shift estimate, not the 0.252 mean difference or the 0.286 observed median difference. Recomputing the original script verifies numerical reproduction, not independent validation of the scoring model or statistical assumptions.

Control was the sole hypothesis-tested outcome. The component means in Table V describe its pattern rather than constitute separate hypothesis tests. Atomic consistency, split-state outcomes, and rollback timing provide descriptive context. Chain verification was also reported separately, although its treatment traceability gate created an indirect dependency within Control.

TABLE V. CONTROL COMPONENTS AND SELECTED CONTEXTUAL OUTCOMES

| Outcome | Operator-driven baseline | State-machine treatment | Difference or scope |
|---|---|---|---|
| Traceability, M (SD) | 0.472 (0.506) | 0.972 (0.167) | 0.500 |
| Accountability, M (SD) | 0.694 (0.467) | 0.972 (0.167) | 0.278 |

| Outcome | Operator-driven baseline | State-machine treatment | Difference or scope |
|---|---|---|---|
| Audit-chain completeness, M (SD) | 0.797 (0.157) | 0.972 (0.167) | 0.175 |
| Reliable rollback capability, M (SD) | 0.917 (0.280) | 0.972 (0.167) | 0.056 |
| Atomic consistency | 34/36 (94.4%) | 36/36 (100%) | Contextual |
| Machine-verifiable audit chain | 0/36 (0%) | 35/36 (97.2%) | Reported separately; gates treatment traceability |
| Split-state outcomes | 2/36 (5.6%) | 0/36 (0%) | Contextual |
| Successful bounded rollback, timed cases | 1/4 (25%) | 3/4 (75%) | Four cases per condition |
| Median recorded rollback time | 21,015.5 ms | 741 ms | Four cases per condition |

Four convergence scenarios per condition produced corrective-rollback timings. One baseline and three treatment reversions completed within 60,000 ms. In S-036, deliberately degraded connector latency breached the bound in both conditions: 143,720 ms for baseline and 120,561 ms for treatment. The full-sample rollback component also included cases requiring no reversion, which scored 1.

S-024 was the only baseline-favoring pair. The governed procedure rejected an unauthenticated Alert-to-Enforce attempt before side effects but returned without retaining an audit-chain entry. Treatment traceability, accountability, and completeness therefore scored zero, yielding Control = 0.250. This was correct prevention without retained refusal evidence, not unauthorized enforcement.

### 5.2. Exploratory Sensitivity Results

Across the 286 weight vectors, mean treatment-minus-baseline differences ranged from 0.056 to 0.500. Assigning 0.4 respectively to traceability, accountability, completeness, and rollback yielded differences of 0.302, 0.257, 0.237, and 0.213. The original equal-weight mean difference was 0.252; its exploratory 95% percentile resampling interval was [0.170, 0.331]. All component mean differences were positive, but individual failures remained. These descriptive checks concern weighting and resampling of this fixed library; they neither validate the construct nor establish a population confidence interval, scoring equivalence, or production effectiveness.

## 6. Discussion

The governed condition had higher measured Control under matched inputs and the disclosed scoring rules. It changed validation, attribution, recordkeeping, commitment, convergence checking, and recovery together. Because arm-specific measurement also differed, the comparison does not isolate either the causal contribution of each mechanism or a scoring-independent governance effect. Sensitivity to component weights cannot resolve this measurement limitation.

Traceability had the largest recorded component contrast (0.500). Treatment generally retained the actor, evidence, and transition, while baseline records depended on scripted profiles. Because traceability predicates differed between conditions, this contrast combines observed recordkeeping with the operational scoring definitions; it is not an independently identified component effect. S-036 additionally shows that governed recovery cannot guarantee a timely response from an impaired dependency.

### 6.1. Candidate Design Principles

The evaluation motivates five candidate principles: first, separate recommendation from commitment authority; second, check explicit guards before enforcement; third, retain attributable evidence for every attempt, including refusals; fourth, govern rollback with explicit recovery conditions and bounds; and fifth, verify observed convergence before declaring completion. S-024 and S-036 expose failures relevant to principles 3 and 4. These principles synthesize the design and boundary cases; they were not independently manipulated or established as universal causal laws.

### 6.2. Operational Implications

For practice, the findings motivate a governance layer between AI recommendations and enforcement authority. Each proposed change should specify identity, evidence, permitted movement, guards, postconditions, convergence criteria, and a rollback target. A confidence threshold alone should not authorize commitment. These are recommendations for further validation, not demonstrated production guarantees.

Refused attempts should be recorded as well as committed transitions. S-024 shows why correct refusal alone does not satisfy the evidence requirements. Production designs should preserve attributable, tamper-evident records of attempted transitions, supporting evidence, guard decisions, and outcomes. The simulated evaluation did not establish persistence across process failure or independently protected audit storage.

Adoption should proceed through observation, approval-gated execution, and only then possible automation after environment-specific testing. Each production adapter requires validation of API versions, permissions, rate limits, compilation delays, and failure behavior. Operational use of the aggregate Control Score also requires independent measurement validation.

For security operations centers, attributable state records could support forensic reconstruction and operator review. For audit and compliance teams, they could support evidence preservation and policy oversight, without establishing regulatory compliance. Incident response could benefit from explicit recovery targets. Production studies must test concurrent updates, administrative overrides, asynchronous propagation, authentication failures, API limits, and large policy sets; deterministic failure cases here do not establish behavior at operational scale.

### 6.3. Limitations

The evaluation used a purposive, constructed scenario library rather than a probability sample of production transitions. The observed effect is therefore conditional on the 36 scenarios and should not be treated as a population estimate for all Zero Trust environments. Although CICIDS2017 and CICIoT2023 supplied real intrusion telemetry records, 14 scenarios were synthetic because authentication failure, connector failure, and rollback-bound conditions were not directly available in those datasets.

The operator-driven baseline consisted of scripted cautious, hasty, and nominal profiles. This design improved determinism and within-pair comparability, but it did not measure human analyst performance, fatigue, expertise, collaboration, or adaptation. The comparison is between the implemented procedures, not between the artifact and all cybersecurity professionals.

Check Point Smart-1 and Palo Alto Panorama behavior was represented through deterministic connector simulations. Production control planes may differ in API behavior, authentication, rate limiting, policy compilation, latency, and transient failure. Consequently, the study demonstrates controlled artifact utility rather than production reliability or readiness.

Control is a study-specific operational measure, not an externally validated instrument. Traceability, accountability, and completeness can count overlapping recordkeeping behavior; arm-specific predicates also limit comparability. The exploratory weighting analysis shows the direction of the library-level mean is not confined to equal weights, but cannot establish component independence or operationally appropriate weights. Bootstrap resampling assumes exchangeable pairs, whereas these purposively constructed scenarios may share dependencies; its interval describes conditional stability, not population coverage. Rounding limits numerical precision. Wilcoxon symmetry was not separately established; the bootstrap does not repair that assumption or turn the primary test into a population claim [19].

During manuscript revision, post hoc diagnostic replays loaded the retained scenario library and settings. The current implementation reproduced all 36 retained paired result rows. Removing the treatment traceability chain gate, and separately applying the baseline's both-states-recorded criterion to treatment, left every result row unchanged. These checks demonstrate stability under those two traceability changes within this library, not equivalence of all scoring rules or independent observations. They were not part of the original analysis; other component predicates were not harmonized. Retained hashes establish file identity, and replay agreement does not identify the exact historical source revision or establish a historical replay-success rate.

Finally, S-024 and S-036 identify concrete boundaries. The former shows that fail-closed prevention must also record refused attempts. The latter shows that formal rollback can expose nonconvergence without forcing an impaired dependency to recover within a target. Retaining both cases prevents the positive overall finding from being interpreted as complete control in every tested condition.

## 7. Conclusion

Across the 36 matched scenarios, measured Control under the disclosed arm-specific scoring rules averaged 0.972 for the governed condition and 0.720 for the baseline. The primary comparison yielded $V = 359$, $p < .001$, and rank-biserial $r = .899$. Twenty-six pairs favored treatment, nine were tied, and one favored baseline. These are results for the implemented conditions, not an isolated estimate of governance effectiveness independent of measurement design.

The contribution is a defined transition-governance procedure and evidence of utility under controlled conditions. It is not a guarantee of production readiness, exhaustive correctness, or a universally validated Control measure. The findings remain conditional on the scenario library, operator model, simulated connectors, and disclosed scoring rules. Conceptually, the study positions transition governance as a candidate architectural layer between recommendation and enforcement, integrating existing theories rather than establishing a new general theory.

Live-analyst studies, independent comparator implementations, production-representative connectors, concurrent transitions, and additional recovery trials remain unperformed extensions. Four rollback-timing cases per condition cannot establish general recovery reliability. These gaps require new evidence, not stronger wording; their absence limits the current contribution to the controlled artifact and retained-data analyses.

### Conflict of Interest

The authors declare that there are no conflicts of interest regarding the publication of this paper.